# Coverage-dependent magnetic and electronic properties of graphene with Co adatoms

Min Gao[1], Jun Hu[1,2,*]

[1]School of Physical Science and Technology & Jiangsu Key Laboratory of Thin Films, Soochow University, Suzhou 215006, China.

[2]School of Physical Science and Technology, Ningbo University, Ningbo 315211, China.

Decorating two-dimensional materials with transition-metal adatoms is an effective way to bring about new physical properties that are intriguing for applications in electronics and spintronics devices. Here, we systematically studied the coverage-dependent magnetic and electronic properties of graphene decorated by Co adatoms, based on first-principles calculations. We found that the if the Co coverage is larger than 1/3 ML, the Co atoms will aggregate to form a Co monolayer and then a van der Waals bilayer system between the Co monolayer and graphene forms. When the Co coverage is $\lesssim$ 1/3 ML, the Co adatom is spin polarized with spin moment varying from 1.1 ~ 1.4 $\mu_B$. The $d_{xz/yz}$ and $d_{xy/x^2-y^2}$ orbitals of Co hybridize significantly with the π bands of graphene, which generates a series of new bands in the energy range from -2 eV to 1 eV with respect to Dirac point of graphene. In most cases, the new bands near the Fermi level lead to topological states characterized by the quantum anomalous Hall effect.

Key words: graphene, adatom, coverage-dependent, band structure, topological state

[*] Corresponding author. Email: hujun2@nbu.edu.cn

## I. Introduction

Low-dimensional materials are desired due to the accelerating miniaturization of electronic devices in the last two decades. A lot of low-dimensional materials have been discovered and synthesized, including zero-dimensional clusters [1, 2], one-dimensional nanotubes and nanowires [3, 4, 5, 6], atomically thin two-dimensional (2D) materials [7, 8, 9, 10, 11]. Among these low-dimensional materials, graphene has evoked particular interest of researchers in the community of condensed matter physics and materials science, ever since its discovery, because it is a perfect platform to exploring various new physical properties which are promising in miscellaneous applications in electronic and optoelectronic devices [11]. For instance, monolayer graphene was predicted as the first candidate of topological insulators (TIs) which exhibit intriguing quantum spin Hall (QSH) state [12, 13]; the band gap of bilayer and trilayer graphene is tunable depending on the stacking pattern or external electric field [14, 15, 16, 17]; twisted bilayer graphene with magic angle is even superconductor [18, 19].

It is remarkable that a rich diversity of electronic and magnetic properties may be achieved by decorating graphene with adatoms. Hydrogenization and fluorization of graphene turn graphene into new semiconductors graphane and fluorographene, respectively [20, 21]. In, Tl or Au adatoms may enhance the TI gap of graphene at the Dirac cone significantly [22, 23]. The 5d orbitals of some late 5d adatoms strongly hybridize with the $\pi$ bands of graphene, leading to giant TI gap [24]. Graphene with Fe or W adatoms becomes Chern insulator which is characterized by quantum anomalous Hall effect (QAHE) [25, 26]. Interestingly, Co or Rh adatoms on graphene could even results in Chern half metal state which also exhibits the QAHE but only in one spin channel, i.e. one spin channel is Chern insulator while the other spin channel is metal [27]. Magnetic transition-metal adatoms or molecules on graphene may induce Kondo effect, where the local spin moment scatters the $\pi$ electrons of graphene [28, 29, 30, 31]. Some magnetic transition-metal adatoms and dimers on graphene possess giant magnetic anisotropy [32, 33]. Clearly, the physical properties of graphene are closely sensitive to the adatoms. Nevertheless, the coverage of adatoms is relatively small in most studies, so that the direct interaction between the adatoms is ignorable. It is interesting how the adatoms interact with each other if the coverage is high and what are the corresponding physical properties.

In this paper, we chose Co adatom as a prototype to study coverage-dependent magnetic and electronic properties of graphene with Co adatoms (notated as Co/Gr), based on first-principles calculations. We found that the if the Co coverage is larger than 1/3 ML, the Co atoms will aggregate to form a Co monolayer which couples with graphene through weak van der Waals (VDW) interaction. When the Co coverage is $\leqslant$ 1/3 ML, the Co adatom is spin polarized with spin moment varying from 1.1 ~ 1.4 $\mu_B$. The hybridization between the Co adatom and graphene generates a series of new bands. In most cases, the new bands near the Fermi level lead to topological phases.

## II. Computational details

The atomic structures, electronic and magnetic properties were calculated by using the Vienna *ab-initio* simulation package which implements the density functional theory to simulate quantum systems [34, 35]. The interaction between valence electrons and ionic cores was described within the framework of the projector augmented wave method [36, 37]. The energy cutoff for the plane wave basis expansion was 500 eV. The spin-polarized generalized gradient approximation was used for the exchange-correlation potentials [38]. To model different coverage of the Co adatoms, we built a series of supercells ranging from 1×1 to 10×10, and then place one Co adatom in each supercell (see Fig. 1 for some cases). Note that the so called 1×1 supercell is actually the unit cell of graphene. The corresponding coverage of Co varies from 1 monolayer (ML) to 1/100 ML. The Brillouin zone was then sampled by a k-grid mesh of $100/n \times 100/n$ where *n* is the size of the supercell. The atomic positions were fully relaxed using the conjugated gradient method until the force on each atom is smaller than 0.01 eV/Å.

## III. Results and discussion

There are three possible sites for the Co adatom on graphene: hollow, atop and bridge sites [27]. We found that the Co adatom prefers the hollow site at which the energy is smaller by about 0.4 eV than the other two sites for all supercells. When the Co coverage is 1 ML, the Co atoms form a closely packed monolayer with the nearest Co-Co distance of 2.47 Å which is the lattice constant of graphene. The Co monolayer and graphene do not bind together with chemical bonds, instead they interact with each other through the weak VDW force, and the distance between the Co monolayer and graphene is about 3.6 Å, as shown in Fig. 1a. For the other Co coverages, the smallest nearest Co-Co distance is 4.27 Å in the $\sqrt{3} \times \sqrt{3}$ supercell of which the Co coverage is 1/3 ML (see Fig. 1b), so the direct interaction between the Co adatoms is much smaller than that in the Co monolayer. The height of the Co adatoms over the graphene decreases significantly to about 1.58 Å. Consequently, the Co atom bonds to six C atoms, and the Co-C bond lengths are 2.13 Å and almost remain the same for all supercells except the 1×1 unit cell, as shown in Fig. 1b and 1c for the $\sqrt{3} \times \sqrt{3}$ and 2×2 supercells.

The stability of the Co adatom on graphene can be estimated by the adsorption energy ($E_A$) which can be expressed as

$$E_A = E(Gr) + E(Co) - E(Co/Gr), \qquad (1)$$

where E(Gr), E(Co), E(Co/Gr) are the total energies of perfect graphene, an isolate Co atom, and Co/Gr, respectively. The calculated $E_A$ as a function of supercell size is plotted in Fig. 2a. It can be seen that $E_A$ first decreases significantly and reaches the lowest value at 3×3 supercell, and then it increases slightly. The graphene with 1 ML Co (i.e. one Co adatom on each 1×1 graphene unit cell) has the largest $E_A$ of 3.64 eV, while the magnitudes of $E_A$ for the other cases are much smaller and varies from 1.05 eV to 1.30 eV. In fact, for the Co coverage of 1 ML, the adsorption energy mainly originates from the interaction between the Co adatoms. The large $E_A$ for this case implies that the interaction between the Co adatoms are much stronger than that between the Co atom and graphene. Therefore, if we expect to investigate the effect induced by the interaction between the Co adatom and graphene, the coverage of Co cannot be as high as 1 ML to avoid the direct interaction between the Co adatoms. Actually, in a $\sqrt{3}\times\sqrt{3}$ supercell of which the Co coverage is 1/3 ML, the strong direct interaction between the Co adatoms is already weakened drastically, which can be manifested by the moderate $E_A$ of 1.30 eV per Co. Obviously, this is the second highest Co coverage for the uniformly distributed Co adatoms on graphene, as seen in Fig. 1. Note that only uniform distribution of the Co adatoms is considered in this work, but in realistic systems the distribution should be random to some extent. As a consequence, the actual highest coverage of Co to avoid direct interaction between the Co adatoms should be smaller than 1/3 ML. Otherwise, clustering of the Co adatoms will occur. Furthermore, the Co adatoms might diffuse on graphene, so that the Co atoms meet each other and then the strong Co-Co bonds form. Finally, a Co monolayer may be achieved after all the Co adatoms aggregate together. Fortunately, the energy barrier of the Co adatom diffusion on graphene is about 0.4 eV [40], which implies that the aggregation of Co adatoms can be avoided below 100 K. In fact, a recent experiment revealed that over 90% of Co adatoms on graphene exist as monomer [41]. Therefore, the Co adatoms will be bounded on graphene as long as they are separate in the beginning, so that the isolate distribution of the Co adatoms can be achieved.

When a Co atom is adsorbed on graphene, electron charge transfer occurs between the Co adatom and graphene. We calculated the charge transfer based on the Bader charge division scheme [39]. For the case of 1 ML Co, the calculated Bader charge ($Q_B$) is 0.03 electron, which indicates that 0.03 electron per unit cell is donated by graphene to the Co monolayer. The charge transfer is very small, because there is no chemical bond between the Co monolayer and graphene. The charge transfer mainly induced by the different work functions of the Co monolayer and graphene. For the cases with lower coverages, the charge transfer is much more significant, and the Co adatom donates electron charge to graphene. The $Q_B$ of the Co adatom is −0.51 electron at the Co coverage of 1/3 ML and decreases to −0.72 electron at the Co coverage of 0.01 ML. Moreover, the $Q_B$ of the Co adatom changes little when the Co coverage is smaller than 1/16 ML (i.e. the supercell is larger than 4×4).

The Co adatoms at all the considered coverages are spin polarized. In the Co monolayer, the total spin moment ($M_T$) is 1.91 $\mu_B$ per Co atom. In this case, the spin moment is completely contributed from the Co atoms, so the local spin moment on each Co adatom ($M_{Co}$) is also 1.91 $\mu_B$. When the Co coverage decreases (i.e. the supercell size increases), the $M_T$ first decreases significantly and reaches the minimum at the supercell of 3×3 (i.e. 1.01 $\mu_B$ per Co atom), and then increases slowly and almost linearly, as seen in Fig. 2b. Moreover, the $M_{Co}$ at each supercell is larger than the corresponding $M_T$, because the C atoms around the Co adatom are spin-polarized and the local spin moment on the C atoms are opposite to that on the Co adatoms, as shown by the inset in Fig. 2b. Obviously, with the Co coverages of 1/3 ML and 0.25 ML (i.e. $\sqrt{3}\times\sqrt{3}$ and 2×2 supercells), the values of $M_T$ are larger than the other cases except the Co monolayer. So the interaction between the Co adatoms still exists and is sizable, and it should be mediated by graphene.

Then we investigate the electronic properties of Co/Gr. Figure 3 plots the band structures of 1×1 graphene with one ML Co adatoms. Because there is no notable interaction between the Co monolayer and graphene, the bands belonging to the Co monolayer and graphene can be separated from each other by analyzing the coefficients of the corresponding wavefunctions. As shown in Fig. 3, the bands belonging to graphene keep the same as the isolate graphene. Moreover, it can be seen that the Dirac point of graphene is slightly higher than the Fermi level, because graphene donates a little part of electron charge to the Co monolayer. The bands belonging to the Co monolayer mainly locate between −3 eV to 1 eV relative to the Fermi level, and are clearly spin-polarized. Furthermore, the spin-orbit coupling (SOC) does not affect the band structure in this case.

For the Co coverage smaller than 1 ML, the band structures are much different from the case of 1 ML Co adatoms, because the direct interaction between the Co adatoms disappears. Instead, the $3d$ orbitals of Co hybridize with the $\pi$ bands of graphene strongly, which generates a series of new bands [24]. Figure 4 plots the band structures of graphene with the Co adatoms at the coverages of 1/3, 1/4, 1/16, 1/64 MLs. It can be seen that the dispersions of the bands depend on the coverage of Co. Note that for the $\sqrt{3}\times\sqrt{3}$ and 3×3 supercells, the K point of the Brillouin zone is folded onto the $\Gamma$ point, so that the Dirac point belonging to graphene shifts to the $\Gamma$ point, as shown in Fig. 4(a/b) and 4(e/f). For the other cases, the Dirac point still locates at the K point. Furthermore, the spin polarization induced by the Co adatoms splits the Dirac bands, which makes the Dirac points in the majority- and minority-spin channels separate from each other, as shown in Fig. 4(c/d) and (g-j).

The hybridization between the $3d$ orbitals of Co and the $\pi$ bands of graphene generates a few new bands, and these bands become more and more localized as the Co coverage decreases. In the $\sqrt{3}\times\sqrt{3}$ and 2×2 supercells, all the orbitals of the Co adatom hybridize with graphene, so almost all the bands in the energy range from -2 eV to 1 eV are contributed from the Co atoms significantly, as seen in Fig. 4b and 4d where the fat bands indicate the weights of the Co atoms. In the 3×3 supercell, the $d_{z^2}$ orbital almost does not hybridize with graphene, because the corresponding bands are quite flat (see Fig. 4e). The bands from the $s$ orbital are dispersive slightly, implying weak hybridization between it and graphene. When the supercell reaches 4×4 (i.e. the Co coverage $\geqslant$ 1/16 ML), both the $s$ and $d_{z^2}$ orbitals remain as atomic orbitals, which results in flat bands in the band structure (Fig. 4g and 4i). The bands contributed from the $d_{xz/yz}$ and $d_{xy/x^2-y^2}$ orbitals are also localized near the Γ point, although they still hybridize with graphene, as shown in Fig. 4g and 4i. In addition, the dispersions of the band structures are similar for the cases with the Co coverage ⩾ 1/16 ML.

It is known that the spin-orbit coupling (SOC) effect is crucial for the topological phases [12, 13]. Accordingly, we calculated the band structures of Co/Gr with the SOC effect. First of all, for the 3n×3n supercells (n is positive integer), Co/Gr is topologically trivial, because the Dirac point shifts to the Γ point and the scattering between the Dirac states and the states at the initial Γ point eliminates the topological feature of the Dirac states [22]. For the supercell reaches 4×4, the degenerate bands of the $d_{xz/yz}$ orbital at the Γ point near the Fermi level split induced by the SOC effect, as seen in Fig. 4f and 4h. The splitting strength only vary slightly in a wide range of the Co coverage, from 58 meV for the Co coverage of 1/16 ML to 50 meV for the Co coverage of 1/100 ML. In fact, the states from the $d_{xz/yz}$ orbital at the Γ point preserve the atomic feature, so the splitting strength is closely associated with the SOC strength of the Co-3d orbitals, rather than the periodict Bloch states of the graphene. Interestingly, this SOC induced splitting makes the minority-spin channel become a semiconductor with the band gap the same as the splitting strength, since the bands crossing the Fermi level around the K point are contributed completely from the majority-spin electrons. As a consequence, Co/Gr becomes a half metal for the Co coverage ⩾ 1/16 ML. Furthermore, the minority-spin channel is a Chern insulator which exhibits the QAHE, so Co/Gr with the Co coverage ⩾ 1/16 ML can be called as Chern half metal [27]. We should point out that the topological phase should retain for the Co coverage much smaller than 1/100 ML. However, we did not consider this in the present work because the calculation requires too much computational resources.

It is interesting to see whether the topological phase exists in the cases with high Co coverage. Let us first investigate the $\sqrt{3}\times\sqrt{3}$ supercell. As plotted in Fig. 4a, the dispersion characteristic of the minority-spin bands at the K point near the Fermi level is very similar with the Dirac bands of pure graphene, so these bands can be regarded as new Dirac bands. This is intriguing because the Dirac bands of pure graphene has been folded around the Γ point in the Brillouin zone of the $\sqrt{3}\times\sqrt{3}$ supercell. In fact, these bands are mostly contributed from the Co adatom, as seen in Fig. 4b. By analyzing the wavefunctions of these new Dirac bands, we found that all the components except $d_{z^2}$ of the Co-3d orbital contribute to the Dirac point. Although there is no direct interaction between the Co adatoms in the $\sqrt{3}\times\sqrt{3}$ supercell, the Co adatoms still interact with each other and the interaction is mediated by graphene. Accordingly, there are a series of dispersive bands contributed from the Co adatom (Fig. 4b). Interestingly, the SOC effect opens a sizable gap (44 meV) at the new Dirac point, and the Fermi level locates in this gap. As a result, Co/Gr with the Co coverage of 1/3 ML is also a half metal. Since the gap in the minority-spin channel is opened by the SOC effect, it should be topologically nontrivial similar to the cases with the Co coverage ⩾ 1/16 ML. In other words, Co/Gr with the Co coverage of 1/3 ML is also a Chern half metal.

For the 2×2 supercell of which the Co coverage is 1/4 ML, the Dirac point of graphene in the majority-spin channel locates slightly above the Fermi level, which leaves a small hole pocket in the band structure. Meanwhile, the minority- and majority-spin bands cross each other near the middle points along the paths $\overline{\Gamma M}$ and $\overline{\Gamma K}$. The degeneracy at the Dirac point and the crossing points are removed by the SOC effect, which opens gaps ~ 12 meV. However, the gap at the Dirac point locates above the Fermi level, while the other two gaps locate bellow the Fermi level. Therefore, Co/Gr in this case is metallic.

It is well known that applying biaxial strain is an effective way to tune the band structures of 2D materials [42, 43]. Accordingly, we applied biaxial strain onto Co/Gr with the Co coverage of 1/4 ML. On the other hand, applying extensile strain onto 2D materials is much easier than applying compressive strain, so we considered biaxial extensile strain from 1% to 10%, with increment of 1%. Figure 5 plots the band structures at the biaxial strains of 2%, 4%, 6% and 8%. Obviously, the Dirac point in the majority-spin channel at the K point is affected by the strain significantly, shifts downwards as the biaxial strain increases. At the biaxial strains of 6%, the Dirac point locates at the Fermi level (without the SOC effect). For larger biaxial strains, the Dirac point are below the Fermi level. For the other two crossing points (denoted by the blue circles in Fig. 5a), they shift upwards slightly and towards the Γ point as the biaxial strain increases. Interestingly, these crossing points also match the Fermi level at the biaxial strain of 6%. As a result, when the SOC effect is considered, gaps of 9, 16 and 11 meV are opened at these points, as shown in Fig. 4f. Similarly, these gaps should be topologically nontrivial since these gaps are induced by the SOC effect. Meanwhile, Co/Gr with the Co coverage of 1/4 ML is spin polarized, so it is a Chern insulator and the QAHE can be observed in this system.

**IV. Conclusions**

In summary, first-principles calculations were carried out to study the stability, magnetic and electronic properties of graphene with Co adatoms. We found that if the Co coverage is 1 ML, the Co atoms form a metallic layer which interact with graphene through weak van der Waals interaction. The formation of the Co monolayer may be avoided if the Co coverage is ≲ 1/3 ML. Then the Co adatom is spin polarized with spin moments varying from 1.1 ~ 1.4 $\mu_B$. When the Co coverage is 1/3 ~ 1/4 ML, the 4s orbital and all the components of the 3d orbital of Co hybridize with graphene, while for the lower coverages, only the $d_{xz/yz}$ and $d_{xy/x^2-y^2}$ components hybridize significantly with graphene. The hybridization generates a series of new bands, and the bandwidth becomes smaller and smaller as the Co coverage decreases. For the Co coverage of 1/(3n×3n) i.e. one Co adatom in a 3n×3n supercell, Co/Gr is topologically trivial. For the other cases, when the Co coverage is 1/3 ML or ≲ 1/16 ML, Co/Gr is Chern half metal in which the QAHE can be observed in one spin channel and the other spin channel is metallic. When the Co coverage is 1/4 ML, the biaxial extensile strain of 6% may turn it into a Chern insulator.

**Acknowledgements**

This work is supported by the National Natural Science Foundation of China (11574223), the Six Talent Peaks Project of Jiangsu Province (2019-XCL-081), and the start-up funding of Ningbo University.

**Reference**
1. E. A. Rohlfing, D. M. Cox, and A. Kaldor, Production and characterization of supersonic carbon cluster beams. J. Chem. Phys. 81, 3322 (1984).
2. T. G. Schmalz, W. A. Seitz, D. J. Klein, and G. E. Hite, Elemental Carbon Cages. J. Am. Chem. Soc. 110, 1113 (1988).
3. S. Iijima, Helical microtubules of graphitic carbon. Nature 354, 56 (1991).
4. N. Hamada, S. Sawada, and A. Oshiyama, New one-dimensional conductors: graphitic microtubules. Phys. Rev. Lett. 68, 1579 (1992).
5. T. M. Whitney, P. C. Searson, J. S. Jiang, and C. L. Chien, Fabrication and Magnetic Properties of Arrays of Metallic Nanowires. Science 261, 1316 (1993).
6. Y. N. Xia, P. D. Yang, Y. G. Sun, Y. Y. Wu, B. Mayers, B. Gates, Y. D. Yin, F. Kim, and H. Q. Yan, One-dimensional nanostructures: synthesis, characterization, and applications. Adv. Mater. 15, 353 (2003).
7. K. S. Novoselov, A. K. Geim, S. V. Morozov, D. Jiang, Y. Zhang, S. V. Dubonos, I. V. Grigorieva, and A. A. Firsov, Electric field effect in atomically thin carbon films, Science 306, 666 (2004).
8. K. S. Novoselov, D. Jiang, F. Schedin, T. J. Booth, V. V. Khotkevich, S. V. Morozov, and A. K. Geim, Two-dimensional atomic crystals. Proc. Natl. Acad. Sci. 102, 10451 (2005).
9. K. F. Mak, C. Lee, J. Hone, J. Shan, and T. F. Heinz, Atomically Thin $MoS_2$: A New Direct-Gap Semiconductor. Phys. Rev. Lett. 105, 136805 (2010).
10. K. K. Liu, W. Zhang, Y. H. Lee, Y. C. Lin, M. T. Chang, C. Y. Su, C. S. Chang, H. Li, Y. Shi, H. Zhang, C. S. Lai, and L. J. Li, Growth of Large-Area and Highly Crystalline $MoS_2$ Thin Layers on Insulating Substrates. Nano Lett. 12, 1538 (2012).
11. X. M. Li, L. Tao, Z. F. Chen, H. Fang, X. S. Li, X. R. Wang, J.-B. Xu, and H. W. Zhu, Graphene and related two-dimensional materials: Structure-property relationships for electronics and optoelectronics, Appl. Phys. Rev. 4, 021306 (2017).
12. C. L. Kane and E. J. Mele, $Z_2$ Topological Order and the Quantum Spin Hall Effect. Phys. Rev. Lett. 95, 146802 (2005).
13. C. L. Kane and E. J. Mele, Quantum Spin Hall Effect in Graphene. Phys. Rev. Lett. 95, 226801 (2005).

14. T. Ohta, A. Bostwick, T. Seyller, K. Horn, and E. Rotenberg, Controlling the Electronic Structure of Bilayer Graphene. Science 313, 951 (2006).
15. E. McCann and M. Koshino, The electronic properties of bilayer graphene. Rep. Prog. Phys. 76, 056503 (2013).
16. C. H. Lui, Z. Li, K. F. Mak, E. Cappelluti, and T. F. Heinz, Observation of an electrically tunable band gap in trilayer graphene. Nat. Phys. 7, 944 (2011).
17. W. Bao, L. Jing, J. Velasco Jr, Y. Lee, G. Liu1, D. Tran, B. Standley, M. Aykol, S. B. Cronin, D. Smirnov, M. Koshino, E. McCann, M. Bockrath, and C. N. Lau, Stacking-dependent band gap and quantum transport in trilayer graphene. Nat. Phys. 7, 948 (2011).
18. Y. Cao, V. Fatemi, S. Fang, K. Watanabe, T. Taniguchi, E. Kaxiras, and Pablo Jarillo-Herrero, Unconventional superconductivity in magic-angle graphene superlattices. Nature 556, 43 (2018).
19. M. Yankowitz, S. Chen, H. Polshyn, Y. Zhang, K. Watanabe, T. Taniguchi, D. Graf, A. F. Young, and C. R. Dean, Tuning superconductivity in twisted bilayer graphene. Science 363, 1059 (2019).
20. J. O. Sofo, A. S. Chaudhari, and G. D. Barber, Graphane: A two-dimensional hydrocarbon. Phys. Rev. B 75, 153401 (2007).
21. O. Leenaerts, H. Peelaers, A. D. Hernández-Nieves, B. Partoens, and F. M. Peeters, First-principles investigation of graphene fluoride and graphane. Phys. Rev. B 82, 195436 (2010).
22. C. Weeks, J. Hu, J. Alicea, M. Franz, and R. Wu, Engineering a Robust Quantum Spin Hall State in Graphene via Adatom Deposition. Phys. Rev. X 1, 021001 (2011).
23. D. Van Tuan, J. M. Marmolejo-Tejada, X. Waintal, B. K. Nikolić, S. O. Valenzuela, and S. Roche, Spin Hall Effect and Origins of Nonlocal Resistance in Adatom-Decorated Graphene. Phys. Rev. Lett. 117, 176602 (2016).
24. J. Hu, J. Alicea, R. Wu, and M. Franz, Giant Topological Insulator Gap in Graphene with 5d Adatoms. Phys. Rev. Lett. 109, 266801 (2012).
25. W.-K. Tse, Z. Qiao, Y. Yao, A. H. MacDonald, and Q. Niu, Quantum anomalous Hall effect in single-layer and bilayer graphene. Phys. Rev. B 83, 155447 (2011).
26. H. Zhang, C. Lazo, S. Blügel, S. Heinze, and Y. Mokrousov, Electrically Tunable Quantum Anomalous Hall Effect in Graphene Decorated by 5d Transition-Metal Adatoms. Phys. Rev. Lett. 108, 056802 (2012).
27. J. Hu, Z. Zhu, and R. Wu, Chern Half Metals: A New Class of Topological Materials to Realize the Quantum Anomalous Hall Effect. Nano Lett. 15, 2074 (2015).
28. T. O. Wehling, A. V. Balatsky, M. I. Katsnelson, A. I. Lichtenstein, and A. Rosch, Orbitally controlled Kondo effect of Co adatoms on graphene. Phys. Rev. B 81, 115427 (2010).
29. T. O. Wehling, A. I. Lichtenstein, and M. I. Katsnelson, Transition-metal adatoms on graphene: Influence of local Coulomb interactions on chemical bonding and magnetic moments. Phys. Rev. B 84, 235110 (2011).

30. V. W. Brar, R. Decker, H.-M. Solowan, Y. Wang, L. Maserati, K. T. Chan, H. Lee, C. O. Girit, A. Zettl, S. G. Louie, M. L. Cohen, and M. F. Crommie, Gate-controlled ionization and screening of cobalt adatoms on a graphene surface. Nat. Phys. 7, 43 (2011).
31. B. Yan, X. Li, J. Zhao, Z. Jia, F. Tang, Z. Zhang, D. Yu, K. Liu, L. Zhang, X. Wu, Gate tunable Kondo effect in magnetic molecule decorated graphene. Solid State Commun. 278, 24 (2018).
32. J. Hu and R. Wu, Giant Magnetic Anisotropy of Transition-Metal Dimers on Defected Graphene. Nano Lett. 14, 1853 (2014).
33. R. Baltic, M. Pivetta, F. Donati, C. Wäckerlin, A. Singha, J. Dreiser, S. Rusponi, and H. Brune, Superlattice of Single Atom Magnets on Graphene. Nano Lett. 16, 7610 (2016).
34. G. Kresse and J. Furthmüller, Efficiency of ab-initio total energy calculations for metals and semiconductors using a plane-wave basis set, Comput. Mater. Sci. 6, 15 (1996).
35. G. Kresse and J. Furthmüller, Efficient iterative schemes for ab initio total-energy calculations using a plane-wave basis set, Phys. Rev. B 54, 11169 (1996).
36. P. E. Blöchl, Projector augmented-wave method, Phys. Rev. B 50, 17953 (1994).
37. G. Kresse and D. Joubert, From ultrasoft pseudopotentials to the projector augmented-wave method, Phys. Rev. B 59, 1758 (1999).
38. J. P. Perdew, K. Burke and M. Ernzerho, Generalized gradient approximation made simple, Phys. Rev. Lett. 77, 3865 (1996).
39. W. Tang, E. Sanville, G. Henkelman, A grid-based Bader analysis algorithm without lattice bias. J. Phys.: Condens. Matter 21, 084204 (2009).
40. O. V. Yazyev and A. Pasquarello, Metal adatoms on graphene and hexagonal boron nitride: Towards rational design of self-assembly templates. Phys. Rev. B 82, 045407 (2010).
41. F. Donati, L. Gragnaniello, A. Cavallin, F. D. Natterer, Q. Dubout, M. Pivetta, F. Patthey, J. Dreiser, C. Piamonteze, S. Rusponi, and H. Brune, Tailoring the Magnetism of Co Atoms on Graphene through Substrate Hybridization. Phys. Rev. Lett. 113, 177201 (2014).
42. L. Y. Ouyang, G. Hu, C. Qi, and J. Hu, Alloying-induced topological transition in 2D transition-metal dichalcogenide semiconductors. Appl. Phys. Express 12, 045003 (2019).
43. G. Hu and J. Hu, Topological Transition in Monolayer Blue Phosphorene with Transition-Metal Adatom under Strain. Chin. J. Chem. Phys. 33, 443 (2020).

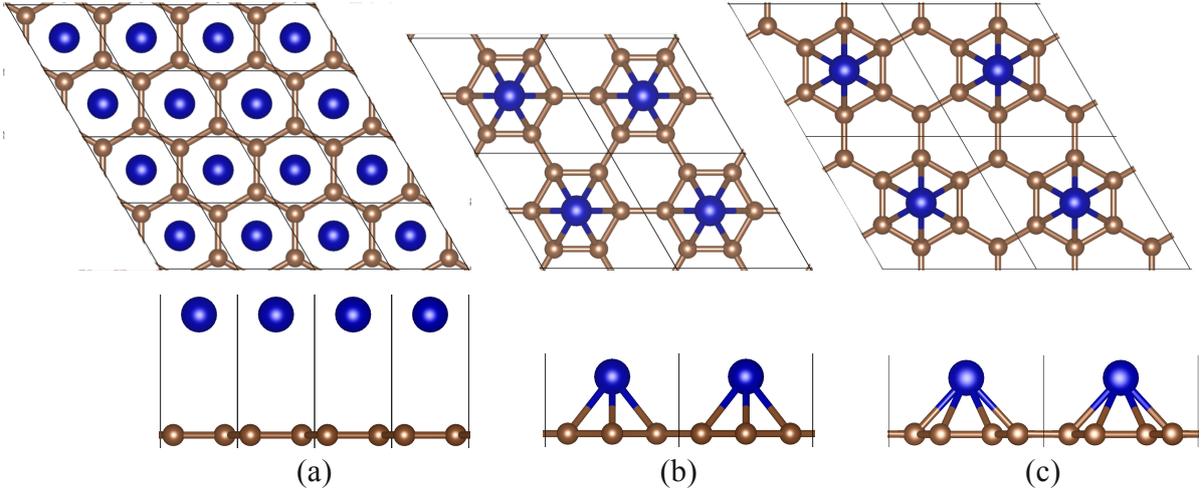

Figure 1. Top and side views of the atomic structure of graphene with Co adatoms with a supercell of (a) 1 × 1, (b) √3 × √3, and (c) 2 × 2. The brown and blue spheres represent the C and Co atoms, respectively. The rhombuses denote the sizes of supercells.

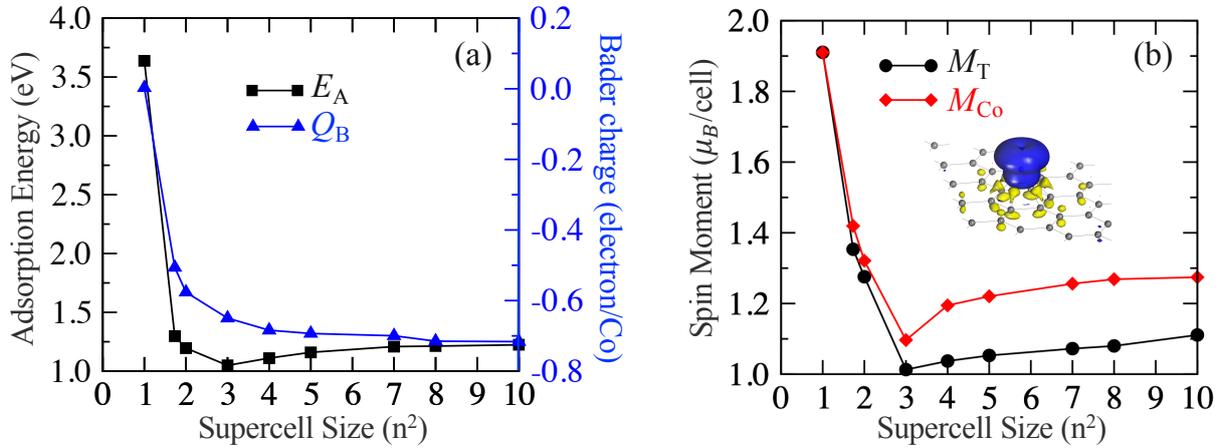

Figure 2. (a) Adsorption energy ($E_A$) and Bader charge ($Q_B$) of Co adatom on graphene with different supercell sizes. (b) Total spin moment ($M_T$) of graphene with Co adatom and local spin moment ($M_{Co}$) on Co. The inset shows the spin density of the Co adatom on a 4×4 graphene with a cutoff of 0.01 $\mu_B/\text{Å}^3$. The blue and yellow isosurfaces indicate the majority- and minority-spin densities, respectively.

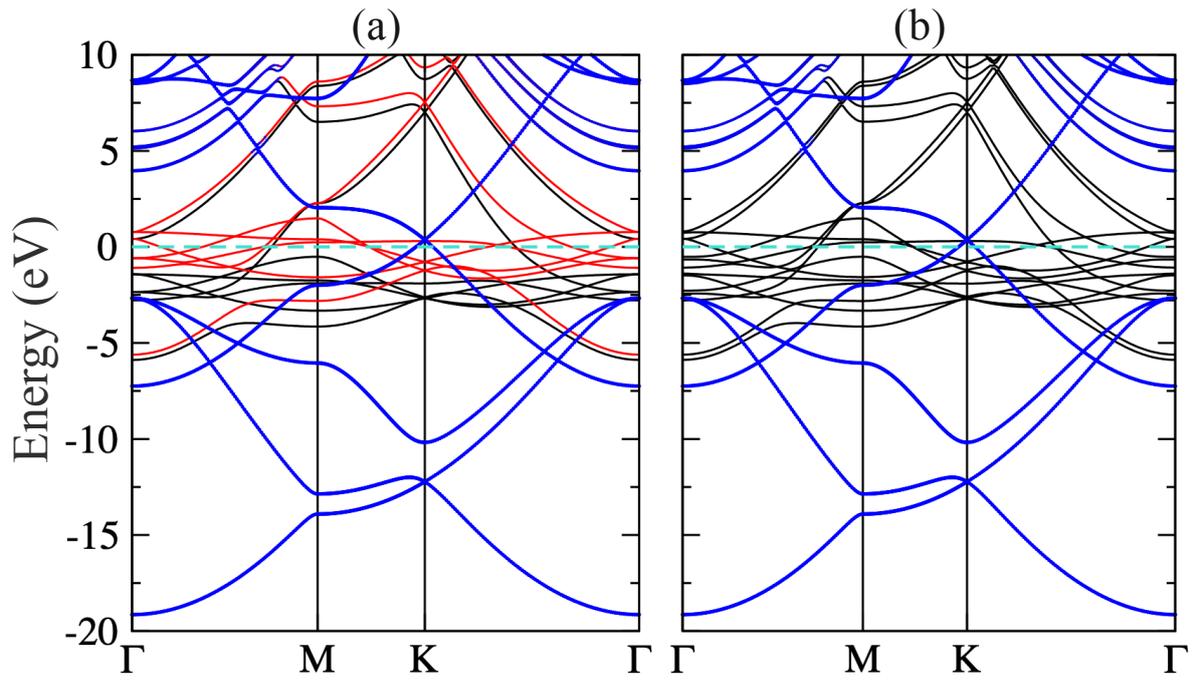

Figure 3. Band structure of graphene with one monolayer Co adatoms (a) without and (b) with the spin-orbit coupling. The black and red curves are for majority- and minority-spin channels, respectively, and they are from the Co adatom. The blue bands are belonging to graphene. The Fermi level (represented by the horizontal dashed lines) is set to zero energy.

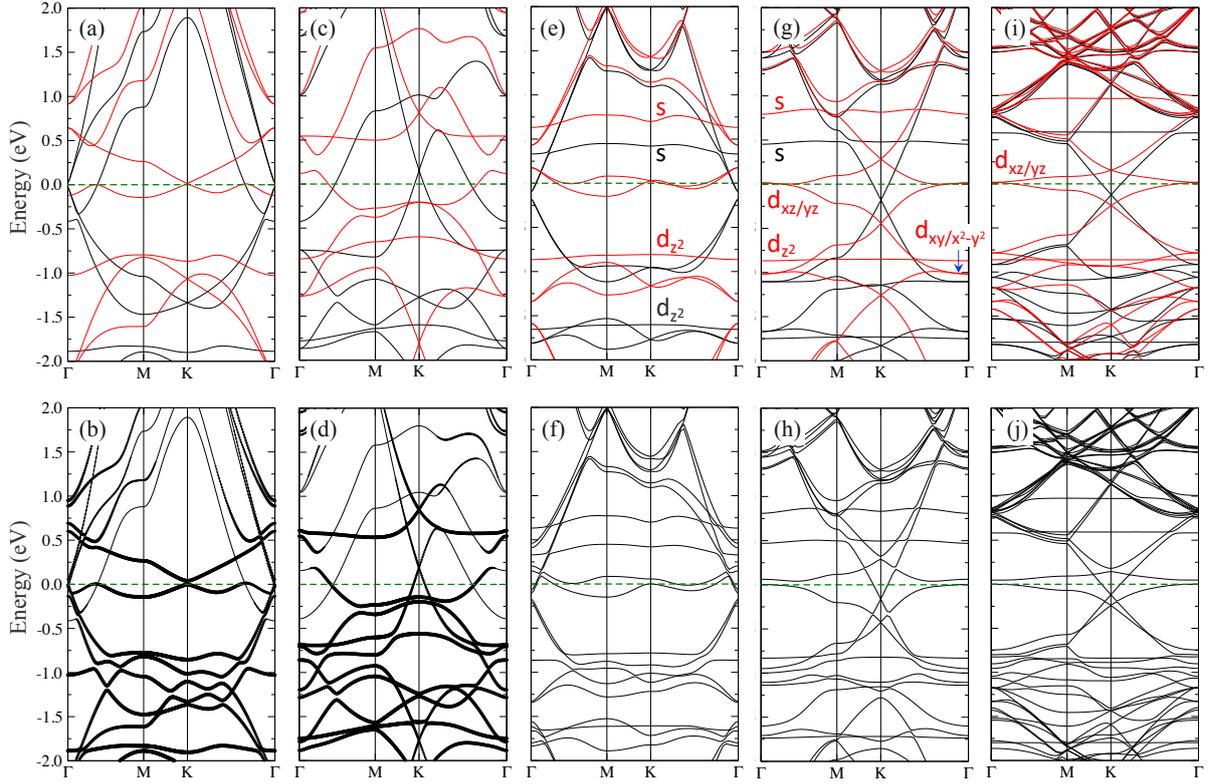

Figure 4. Band structures of graphene with Co adatom without (top panels) and with (bottom panels) the spin-orbit coupling. The supercell sizes are (a/b) $\sqrt{3} \times \sqrt{3}$, (c/d) $2 \times 2$, (e/f) $3 \times 3$, (g/h) $4 \times 4$, and (i/j) $8 \times 8$. The black and red curves in top panels are for majority- and minority-spin channels, respectively. The Fermi level (represented by the horizontal dashed lines) is set to zero energy. (b, d) The thickness of bands indicates the weight of Co atom.

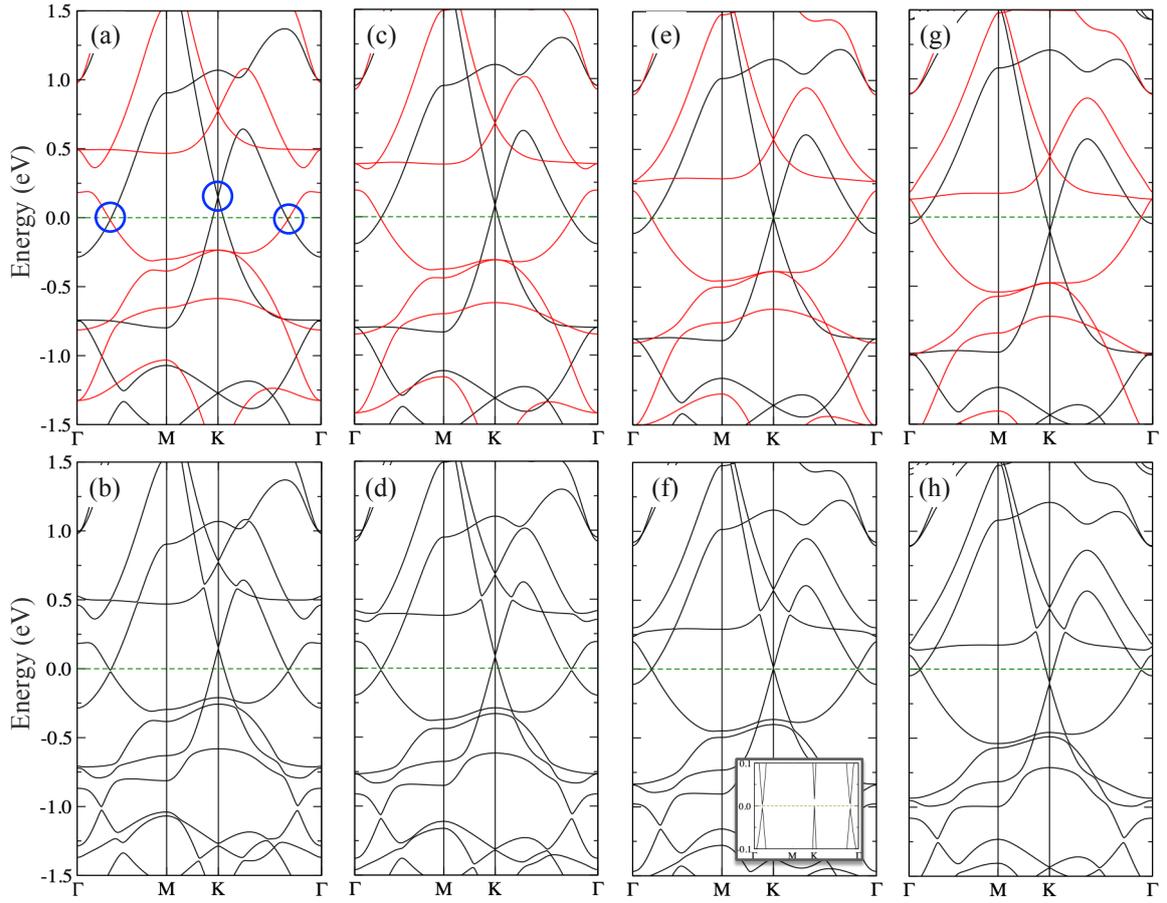

Figure 5. Band structures of graphene with Co coverage of 1/4 ML under biaxial extensile strain without (top panels) and with (bottom panels) the spin-orbit coupling. The biaxial extensile strain is (a/b) 2%, (c/d) 4%, (e/f) 6%, and (g/h) 8%. The black and red curves in top panel are for majority- and minority-spin channels, respectively. The Fermi level (represented by the horizontal dashed lines) is set to zero energy. The blue circles denote the Dirac point in the majority-spin channel and two crossing points near the Fermi level. The inset in (f) zooms in the bands from -0.1 eV to 0.1 eV.